\documentstyle[12pt]{article}
\begin{document}

\catcode`\@=11
\long\def\@makefntext#1{ 
\protect\noindent \hbox to 3.2pt {\hskip-.9pt
$^{{\ninerm\@thefnmark}}$\hfil}#1\hfill} 

\def\thefootnote{\fnsymbol{footnote}}
 \def\@makefnmark{\hbox to 0pt{$^{\@thefnmark}$\hss}}  

\def\ps@myheadings{\let\@mkboth\@gobbletwo
\def\@oddhead{\hbox{} 
\rightmark\hfil\ninerm\thepage}
\def\@oddfoot{}\def\@evenhead{\ninerm\thepage\hfil 
\leftmark\hbox{}}\def\@evenfoot{}
\def\sectionmark##1{}\def\subsectionmark##1{}}

\textwidth 6.0in
\textheight 8.5in
\pagestyle{empty}
\topmargin -0.25truein
\oddsidemargin 0.30truein
\evensidemargin 0.30truein
\raggedbottom
\parindent=20pt
\baselineskip=14pt


\newcommand{\symbolfootnote}{\renewcommand{\thefootnote}
	{\fnsymbol{footnote}}}
\renewcommand{\thefootnote}{\fnsymbol{footnote}}
\newcommand{\alphfootnote}
	{\setcounter{footnote}{0}
	 \renewcommand{\thefootnote}{\sevenrm\alph{footnote}}}

\newcounter{sectionc}\newcounter{subsectionc}\newcounter{subsubsectionc}
\renewcommand{\section}[1] {\vspace{0.6cm}\addtocounter{sectionc}{1}
\setcounter{subsectionc}{0}\setcounter{subsubsectionc}{0}\noindent
	{\bf\thesectionc. #1}\par\vspace{0.4cm}}
\renewcommand{\subsection}[1] {\vspace{0.6cm}\addtocounter{subsectionc}{1}
	\setcounter{subsubsectionc}{0}\noindent
	{\it\thesectionc.\thesubsectionc. #1}\par\vspace{0.4cm}}
\renewcommand{\subsubsection}[1]
   {\vspace{0.6cm}\addtocounter{subsubsectionc}{1}
	\noindent {\rm\thesectionc.\thesubsectionc.\thesubsubsectionc.
	#1}\par\vspace{0.4cm}}
\newcommand{\nonumsection}[1] {\vspace{0.6cm}\noindent{\bf #1}
	\par\vspace{0.4cm}}

\newcounter{appendixc}
\newcounter{subappendixc}[appendixc]
\newcounter{subsubappendixc}[subappendixc]
\renewcommand{\thesubappendixc}{\Alph{appendixc}.\arabic{subappendixc}}
\renewcommand{\thesubsubappendixc}
	{\Alph{appendixc}.\arabic{subappendixc}.\arabic{subsubappendixc}}

\renewcommand{\appendix}[1] {\vspace{0.6cm}
        \refstepcounter{appendixc}
        \setcounter{figure}{0}
        \setcounter{table}{0}
        \setcounter{equation}{0}
        \renewcommand{\thefigure}{\Alph{appendixc}.\arabic{figure}}
        \renewcommand{\thetable}{\Alph{appendixc}.\arabic{table}}
        \renewcommand{\theappendixc}{\Alph{appendixc}}
        \renewcommand{\theequation}{\Alph{appendixc}.\arabic{equation}}
        \noindent{\bf Appendix \theappendixc #1}\par\vspace{0.4cm}}
\newcommand{\subappendix}[1] {\vspace{0.6cm}
        \refstepcounter{subappendixc}
        \noindent{\bf Appendix \thesubappendixc. #1}\par\vspace{0.4cm}}
\newcommand{\subsubappendix}[1] {\vspace{0.6cm}
        \refstepcounter{subsubappendixc}
        \noindent{\it Appendix \thesubsubappendixc. #1}
	\par\vspace{0.4cm}}

\def\abstracts#1{{
	\centering{\begin{minipage}{30pc}\tenrm\baselineskip=12pt\noindent
	\centerline{\tenrm ABSTRACT}\vspace{0.3cm}
	\parindent=0pt #1
	\end{minipage} }\par}}

\newcommand{\bibit}{\it}
\newcommand{\bibbf}{\bf}
\renewenvironment{thebibliography}[1]
	{\begin{list}{\arabic{enumi}.}
	{\usecounter{enumi}\setlength{\parsep}{0pt}
\setlength{\leftmargin 1.25cm}{\rightmargin 0pt}
	 \setlength{\itemsep}{0pt} \settowidth
	{\labelwidth}{#1.}\sloppy}}{\end{list}}

\topsep=0in\parsep=0in\itemsep=0in
\parindent=1.5pc

\newcounter{itemlistc}
\newcounter{romanlistc}
\newcounter{alphlistc}
\newcounter{arabiclistc}
\newenvironment{itemlist}
    	{\setcounter{itemlistc}{0}
	 \begin{list}{$\bullet$}
	{\usecounter{itemlistc}
	 \setlength{\parsep}{0pt}
	 \setlength{\itemsep}{0pt}}}{\end{list}}

\newenvironment{romanlist}
	{\setcounter{romanlistc}{0}
	 \begin{list}{$($\roman{romanlistc}$)$}
	{\usecounter{romanlistc}
	 \setlength{\parsep}{0pt}
	 \setlength{\itemsep}{0pt}}}{\end{list}}

\newenvironment{alphlist}
	{\setcounter{alphlistc}{0}
	 \begin{list}{$($\alph{alphlistc}$)$}
	{\usecounter{alphlistc}
	 \setlength{\parsep}{0pt}
	 \setlength{\itemsep}{0pt}}}{\end{list}}

\newenvironment{arabiclist}
	{\setcounter{arabiclistc}{0}
	 \begin{list}{\arabic{arabiclistc}}
	{\usecounter{arabiclistc}
	 \setlength{\parsep}{0pt}
	 \setlength{\itemsep}{0pt}}}{\end{list}}

\newcommand{\fcaption}[1]{
        \refstepcounter{figure}
        \setbox\@tempboxa = \hbox{\tenrm Fig.~\thefigure. #1}
        \ifdim \wd\@tempboxa > 6in
           {\begin{center}
        \parbox{6in}{\tenrm\baselineskip=12pt Fig.~\thefigure. #1 }
            \end{center}}
        \else
             {\begin{center}
             {\tenrm Fig.~\thefigure. #1}
              \end{center}}
        \fi}

\newcommand{\tcaption}[1]{
        \refstepcounter{table}
        \setbox\@tempboxa = \hbox{\tenrm Table~\thetable. #1}
        \ifdim \wd\@tempboxa > 6in
           {\begin{center}
        \parbox{6in}{\tenrm\baselineskip=12pt Table~\thetable. #1 }
            \end{center}}
        \else
             {\begin{center}
             {\tenrm Table~\thetable. #1}
              \end{center}}
        \fi}

\def\@citex[#1]#2{\if@filesw\immediate\write\@auxout
	{\string\citation{#2}}\fi
\def\@citea{}\@cite{\@for\@citeb:=#2\do
	{\@citea\def\@citea{,}\@ifundefined
	{b@\@citeb}{{\bf ?}\@warning
	{Citation `\@citeb' on page \thepage \space undefined}}
	{\csname b@\@citeb\endcsname}}}{#1}}

\newif\if@cghi
\def\cite{\@cghitrue\@ifnextchar [{\@tempswatrue
	\@citex}{\@tempswafalse\@citex[]}}
\def\citelow{\@cghifalse\@ifnextchar [{\@tempswatrue
	\@citex}{\@tempswafalse\@citex[]}}
\def\@cite#1#2{{$\null^{#1}$\if@tempswa\typeout
	{IJCGA warning: optional citation argument
	ignored: `#2'} \fi}}
\newcommand{\citeup}{\cite}

\def\fnm#1{$^{\mbox{\scriptsize #1}}$}
\def\fnt#1#2{\footnotetext{\kern-.3em
	{$^{\mbox{\sevenrm #1}}$}{#2}}}

\font\twelvebf=cmbx10 scaled\magstep 1
\font\twelverm=cmr10 scaled\magstep 1
\font\twelveit=cmti10 scaled\magstep 1
\font\elevenbfit=cmbxti10 scaled\magstephalf
\font\elevenbf=cmbx10 scaled\magstephalf
\font\elevenrm=cmr10 scaled\magstephalf
\font\elevenit=cmti10 scaled\magstephalf
\font\bfit=cmbxti10
\font\tenbf=cmbx10
\font\tenrm=cmr10
\font\tenit=cmti10
\font\ninebf=cmbx9
\font\ninerm=cmr9
\font\nineit=cmti9
\font\eightbf=cmbx8
\font\eightrm=cmr8
\font\eightit=cmti8


\centerline{\tenbf GRAVITY AND EFFECTIVE FIELD THEORY:}
\baselineskip=22pt
\centerline{\tenbf A TALK FOR PHENOMENOLOGISTS}
\vspace{0.8cm}
\centerline{\tenrm JOHN F. DONOGHUE}
\baselineskip=13pt
\centerline{\tenit Department of Physics and Astronomy,
University of Massachusetts}
\baselineskip=12pt
\centerline{\tenit Amherst, MA  ~01003}
\vspace{0.9cm}
\abstracts{
I briefly summarized some recent work which uses the techniques of
effective field theory to make quantum predictions in general relativity.  In
contrast to conventional expectations, these are in fact well behaved.  The
leading quantum correction to the interaction of two heavy masses is used as
a specific example.}

\vspace{0.8cm}

\twelverm   
\baselineskip=14pt

In this talk, I would like to describe some recent ideas concerning the theory
of gravity which make it seem relatively similar to our theories of the other
interactions when viewed at ordinary energies.  We have come to view all of
our theories as effective theories, valid for some range of energies.  At high
enough energies we expect the interactions to change and new degrees of
freedom to appear, modifying the nature of the theory.  The effects of the
new high energy degrees of freedom are suppressed if one only works at
low energy and these are represented by "nonrenormalizable" interactions.
The techniques of effective field theory are useful in dealing with such
interactions.  Effective field theory allows one to separate the quantum
effects of low energy particles from the as-yet-unknown effects of the high
energy theory.  General relativity fits naturally into the framework of an
effective theory, and this provides a methodology for making quantum
predictions at low energy.

This paper is a short summary of a talk in which I described the use of
gravitational effective field theory to an audience which was primarily
phenomenologists.  The length limitation means that only the general ideas
can be superficially described.  Other papers [1, 2, 3] provide more details.

One of the keys to effective field theory
is that the effects of heavy particles
all appear to be local interactions when viewed at low energy.  The
unknown final theory of gravity will then lead to a series of terms in the
most general local Lagrangian consistent with general covariance.  This
leads in general to an infinite number of terms in the Lagrangian.  The
second key to effective field theory is to order these in powers of factors of
low energy scales divided by high energy scales.  This allows one to ignore
highly suppressed interactions and only deal with a finite set of interactions
relevant to the accuracy that one is working.  In the case of gravity, these
ideas combine to form a derivative expansion for the effective Lagrangian.
The curvatures $R_{\mu \nu \alpha \beta}, R_{\mu \nu}$ and $R$ all
involve two derivatives of the gravitational field.  At low energies these
derivatives become factors of the momenta, which are small.  Thus if we
order the Lagrangian in increasing powers of the curvature, the higher order
terms are increasingly smaller.  Since experimentally we know that the
cosmological constant is negligibly small on ordinary scales, the derivative
expansion starts at order $E^2$,

\begin{equation}
S_{grav} = \int d^4x \sqrt{g} \{ {2 \over K^2} R + c_1 R^2 + c_2
R_{\mu \nu} R^{\mu \nu} + \cdots \}
\end{equation}

\noindent Here the first term has a specific coefficient with $K^2 = 32 \pi
G$ where $G$ is Newton's constant, in order to reproduce Newtonian
gravity in the appropriate limit.  The unknown coefficient $c_i$ are
dimensionless and are experimentally bounded [4] by $c_i < 10^{74}$.
The reason that these bounds are so poor is because these terms have very
little effect at ordinary energies.

While the idea of gravity as a classical effective field theory is widely
accepted, the quantum uses of effective field theory for gravity are  more
novel.  The quantizations of gravity goes back to Feynman [5] and De Witt
[6], but was given its most satisfying form by 't Hooft and Veltman [7].
Here one expands the quantum fields about a smooth background space
time.  One adds a gauge fixing term along with the associated ghost fields,
which are fermionic vector fields in the case of gravity.  Vertices and
propagators can be read off as usual by expanding the Lagrangian in powers
of the quantum field.  The quantization itself is not problematic; where
gravity deviates from other theories is in the nature of the corrections
induced by loop diagrams.  Because of the dimensionful coupling constant
$K \sim \sqrt{G}$, one loop diagrams induce corrections proportional not
to the basic Einstein action $R$, but to the higher order Lagrangians [7, 8]
$R^2$ and $R_{\mu \nu} R^{\mu \nu}$.  Because these corrections are not
finite, the theory is technically called "non renormalizable".  However when
treated as an effective field theory
it is easy to renormalize the theory at any
given order by absorbing the divergence's into renormalized value of the
coefficient in the most general effective Lagrangian [2].

The divergences in the quantum corrections come from the high energy end
of the loop integration, where the theory is not to be trusted.  Only the low
energy portion of the loops is reliable, because this portion uses the correct
low energy interactions.  The low energy quantum effects can be identified
because they are non-local, in contrast to the local effects from high
energy.  In momentum space, nonlocality is most clearly manifested in
nonanalytic behavior, which can never be equivalent to a local Lagrangian.
[Some analytic quantum corrections can also come from low energy, but
these often would not be able to be distinguished from the effects of the
local effective Lagrangian.]

The gravitational interaction of two heavy masses at one loop provides an
example of these ideas.  In momentum space, the interaction is given
schematically [2] by

\begin{equation}
V(q) \sim Gm_1 \left[{1 \over q^2} + {1 \over q^2} \left(
\alpha q^4 + \beta  q^4 \sqrt{{m^2 \over -q^2}} + \gamma
q^4 ln (-q^2) \right) {1 \over q^2} \right]  m_2
\end{equation}

\noindent Here I have suppressed all the Lorentz indices.  The analytic term
$(\alpha)$ contains the coefficients $c_i$ from the effective Lagrangian, as
well as divergent contributions from one-loop integral.  The nonanalytic
terms
($\beta$ and $\gamma$) are finite and calculable from the low energy theory.
When one goes to coordinate space, ${1 \over q^2}$ becomes ${1 \over
r}$, the constant term $( \alpha)$ becomes a delta function at the origin, the
$\sqrt{{m^2 \over -q^2}}$ piece behaves as ${1 \over r^2}$ and $ln (-
q^2)$ behaves as ${1 \over r^3}$.  We see that the nonanalytic terms give
the power law corrections to the Newtonian potential, i.e., the long range
corrections.  With a particular definition of the potential, I find [1,2]

\begin{equation}
V(r) = -{Gm_1 m_2 \over r} \left[ 1 - {G(m_1 + m_2) \over rc^2} -
{127 \over 30\pi^2} {G
\hbar \over r^2c} \right]
\end{equation}

As can be seen from the powers of $\hbar$, the first power modification is a
classical effect, due to the nonlinear nature of classical general relativity,
while the second is a quantum correction.  The magnitude of the quantum
term is unmeasureably small, but the specific number is not as important as
the methodology.  The message is that, by use of the techniques of effective
field theory, the leading quantum correction to the Newtonian potential is a
calculable quantity.

The effective field theory of gravity is a conservative approach in that it
addresses only those features of gravity which we have the right to expect to
be valid at energies which are presently accessible.  It makes no
speculations about the ultimate high energy theory of gravity.  This is
similar to our present view of the Standard Model, for which we have come
to expect modification as low as 1 TeV.  Gravity is expected to remain
unchanged up to the Planck scale.

\vspace{12pt}
\noindent{\bf References}

\begin{enumerate}
\item John F. Donoghue, {\em Phys. Rev. Lett.} {\bf 72}, 2996 (1994).
\item John F. Donoghue, {\em Phys. Rev.} {\bf D50}, 3874 (1994).
\item A longer and more pedagogical presentation is given in the written
version of my lectures at the Summer School on Effective Theories,
Almunecar, Spain, June 1995 gr-qc/9512024.
\item K.S. Stelle, {\em Gen. Rel. Grav.} {\bf 9}, 353 (1978).
\item R.P. Feynman, Acta. Phys. Pol. {\bf 24}, 697 (1963); Caltech lectures
1962-63.
\item B.S. De Witt, {\em Phys. Rev.} {\bf 160}, 1113 (1967); ibid {\bf 162},
1195, 1239 (1967).
\item G. 't Hooft and M. Veltman, {\em Ann. Inst. H. Poincare} {\bf A20}, 69
(1974).
\item M. Geroff and A. Sagnotti, {\em Nucl. Phys.} {\bf B266}, 799 (1986).
\end{enumerate}
\end{document}